\documentclass[aps,prl,twocolumn,showpacs,floatfix,preprintnumbers,amsmath,amssymb,superscriptaddress,reprint]{revtex4-1}

\usepackage[colorlinks,bookmarks=false,citecolor=blue,linkcolor=blue,urlcolor=blue]{hyperref}
\usepackage[all]{hypcap}   

\usepackage{amsmath,amssymb}
\usepackage{graphicx}
\graphicspath{{figures/}{/}}
\usepackage{dcolumn}
\usepackage{bm}
\usepackage[capitalise,compress]{cleveref}
\crefname{section}{Sec.}{Secs.}
\Crefname{section}{Section}{Sections}

\usepackage{verbatim}
\usepackage{color}
\usepackage{soul}
\usepackage{placeins}  
\usepackage{flafter}     

\usepackage{dsfont}

\usepackage{color}

\newcommand{\HIDDEN}[1]{}

\makeatletter
\let\Hy@backout\@gobble
\makeatother

\begin{document}

\title{Faster entanglement driven by quantum resonance in many-body kicked rotors}

\author{Sanku Paul}
\email{sankup005@gmail.com}
\affiliation{Department of Physics
and Complex Systems, S.N. Bose National Centre for Basic
Sciences, Kolkata 700106, India}

\author{J. Bharathi Kannan}
\email{jbharathi.kannan@students.iiserpune.ac.in}
\affiliation{Department of Physics, Indian Institute of Science Education and Research, Pune 411008, India}

\author{M.S. Santhanam}
\email{santh@iiserpune.ac.in}
\affiliation{Department of Physics, Indian Institute of Science Education and Research, Pune 411008, India}


\begin{abstract}
Quantum resonance in the paradigmatic kicked rotor system is a purely quantum effect that ignores the state of underlying classical chaos. In this work, it is shown that quantum resonance leads to superlinear entanglement production. In $N$-interacting kicked rotors set to be at quantum resonance, entanglement growth is super-linear until a crossover timescale $t^*$, beyond which growth slows down to a logarithmic form with superimposed oscillations. By mapping positional interaction to momentum space and analytically assessing the linear entropy, we unravel the mechanism driving these two distinct growth profiles. The analytical results agree with the numerical simulations performed for two- and three-interacting kicked rotors. The late time entanglement oscillation is sensitive to changes in scaled Planck's constant with a high quality factor suitable for high precision measurements. These results are amenable to an experimental realization on atom optics setup.
\end{abstract}

\maketitle

Quantum entanglement captures the degree of non-local correlation between two groups of particles and serves as a crucial resource for quantum technologies, {\it e.g.}, quantum computation \cite{Nielsen_2000,Ladd_2010}, teleportation \cite{Bennett_1993,Braunstein_1998,Bouwmeester_1997,Liu_2024}, and secure communication \cite{GisRibTie02,Ekert_1991}.  In many-body quantum systems, entanglement as quantified by the von-Neumann entropy $S_{\rm vN}$ carries signatures of the quantum phases and the phase transitions \cite{Fazio_2008,Horodecki_2009,Nicolas_2016}. For instance, in ergodic phase, $S_{\rm vN}$ grows linearly with time before saturating \cite{Borgonovi_2016,Luca_2016,Aru01,Miller_1999}. In non-ergodic many-body localized phases, asymptotically, $S_{\rm vN}$ displays an even slower logarithmic growth \cite{Bardarson_2012,Serbyn_2013,Chiara_2006}, as also in ergodic two-body kicked rotor variants \cite{Sanku_2020, AnjBhaSan24}. As entanglement is a useful resource for quantum technologies, faster than linear rate of entanglement generation, {\it e.g.}, superlinear growth rate, will be useful in many practical settings in which quantum effects must be exploited within the coherence time \cite{Briegel_04, Zoller_98}.

In recent years, efforts towards faster entanglement generation has received much attention. In non-Hermitian $PT$-symmetric systems, exceptional points at which the eigenlevels and eigenstates coalease can emerge. When externally driven non-Hermitian qubits are weakly coupled, in a parameteric regime close to an exceptional point, entanglement between the qubits is generated rapidly over timescale much faster than the inverse of coupling strength \cite{LiCheAbb2023,ZhaZhoZuo2024}. Another variant of fast entanglement generation problem appeals to quantum speed limit in the framework of Mandelstamm and Tamm \cite{ManTam1945,ManTam1991,DefCam2017}. In this case, (maximally) entangled GHZ \cite{greenberger2007going} and Werner states \cite{Werner_89} are obtained starting from product states, using optimised Hamiltonians that saturate the quantum speed limit bounds \cite{Cie2023}. 

In the former case, enhanced entanglement generation rate depends on being in the vicinity of an exceptional point, which itself will require a system capable of displaying an exceptional point. In the latter case, a target entangled state is necessary to verify if quantum speed limit lower bound is saturated or not. In both these cases, rate of entanglement generation is not an asymptotic effect and also depends on factors such as the choice of initial states and symmetry of the system.

As the ergodic phase of interacting systems can usually deliver linear entanglement production rate \cite{Luca_2016, Aru01}, a question arises if long-lasting and faster-than-linear entanglement production is possible in Hermitian many-body systems. To study this, $N$-body interacting quantum kicked rotors (QKR) is considered with parameters chosen to be at quantum resonance. Physically, quantum resonance arises when the transition frequency between unperturbed states of the system is matched by the external driving frequency. As a quantum effect, it ignores the nature of underlying classical dynamics, and its one manifestation is the ballistic growth of mean energy, $\langle E \rangle \sim t^2$ \cite{casati1979stochastic, Izrailev_1980,Santhanam_2022,Arcy_2001}. It is then physically reasonable to anticipate that, under quantum resonance conditions, entanglement might also grow at rates faster than linear. This cannot happen in the extensively studied off-resonant QKR (with or without interactions) and its variants due to the emergence of dynamical localization that suppresses the mean energy growth \cite{ChiIzrShe81,Izrailev_1990,Santhanam_2022,Sanku_2016,Sanku_2018}, akin to Anderson localization in position space \cite{Grempel_1984,Anderson_1958,Abrahams_1979}. Both the resonant and off-resonant non-interacting QKR have a large body of applications \cite{AdaTodIke88,Gadway_2013,Wang_2008,Sanku_2023,Monterio_2002,Sumit_2017,Sanku_2019,Santhanam_2022,BolGroCla22,ZhoGon18,DasAnd12,SagBhaKus22}, though quantum resonance in {\sl interacting} QKR remains to be studied.

The main contribution of this work is towards unraveling the super-linear entanglement production in Hermitian many-body quantum system, \emph{i.e.}, $N$-interacting QKR, when the quantum resonance condition is satisfied. While finite number of coupled rotors were studied earlier \cite{Sanku_2020,Arul_2020,Arul_2023,Santhanam_2022}, this work presents first results for $N$-interacting QKR. We show that entanglement displays two distinct growth regimes with a crossover at time $t^*$. For $t < t^*$, entanglement grows super-linearly with time, while for $t > t^*$ a logarithmic growth with a superimposed oscillation is observed. The crossover timescale $t^*$ is inversely dependent on the interaction strength. We explain the distinct entanglement growth profiles by analytically evaluating a relatively simpler indicator of entanglement, the linear entropy. The late time oscillations for $t> t^*$ vanish even for slight deviation from resonance condition and in this sense exhibit a high quality factor. This feature is useful for high precision measurement of driving frequencies.

\textit{System}:
The Hamiltonian of the $N$-interacting QKR is
\begin{equation} \label{eq:Ham}
    H=\sum_{i=1}^{N} \frac{\tau_i p_i^2}{2} + \left[V_{\rm kick} + V_{\rm int}\right] \sum_n \delta(t-n T), 
\end{equation}
where $x_i$ and $p_i$ are the position and momentum of the $i$th rotor with mass $1/\tau_i$. Here, we consider $\tau_i\in \mathbb{Z}$ and $\tau_i/\tau_j \neq 1$. The kicking potential is $V_{\rm kick} = \sum_{i=1}^{N} K_i~\cos(x_i+\Phi_i)$  with kick strength $K_i$ and $\Phi_i\neq 0$ breaks the spatial symmetry of the system. In this work, motivated by experiments, an all-to-all interaction between the rotors of the form $V_{\rm int} = K~\cos\left(\sum_{i=1}^N x_i\right)$ is considered, where $K$ is the interaction strength and $T$ is the time period for the application of kick and interaction potentials: $V_{\rm kick}$ and $V_{\rm int}$. The all-to-all interaction in the position space can be mapped to an all-to-all interaction in the momentum space. In an experiment, interaction only in momentum space can be easily generated using lasers with incommensurate frequencies \cite{Gadway_2013}. To obtain the mapping, one needs to perform a coordinate transformation. For instance, consider two-interacting QKR with kick strength $K_i=0, ~(i=1,2)$ in \cref{eq:Ham} and perform a coordinate transformation $\Theta_1=x_1+x_2, ~\Theta_2=x_1-x_2, ~u=p_1+p_2, ~\text{and}~v=p_1-p_2$. Under this transformation, the interaction in position gets mapped to the momentum space as $\eta uv$, where $\eta=\tau_1 -\tau_2$. An interesting effect emerges: the interaction in momentum space is always present unlike that in the position space which is active only at the time of kicking. Consideration of $K_i=0$ will become clear in the subsequent section. Earlier, many variants of two-interacting QKR have been studied extensively \cite{Borgonovi_1995,Qin_2017,Park_2003,Sanku_2020,Arul_2020,Arul_2023,Santhanam_2022,Mu_2024,Schmidt_2023}.

Quantum dynamics is obtained using the time-evolution operator $\mathcal{U}=(U_1\otimes U_2 \otimes \cdots \otimes U_N) U_{\rm int}$, where 
\begin{equation}
\begin{aligned}
 U_i &= U_i^{\rm free} ~ U_i^{\rm kick}, \\
 &=\exp[-i \tau_i p_i^2 T/2 \hbar_s] ~ \exp[-i K_i \cos (x_i+\theta_i)/\hbar_s]   
\end{aligned}
\end{equation}
is the time-evolution operator of the $i$-th rotor and $\hbar_s$ is the scaled Planck's constant. The interaction appears as $U_{\rm int}= \exp[-i K \cos (\sum_{i=1}^N x_i)/\hbar_s]$. Now, starting with initial state in product form, \emph{i.e.}, $|\psi(0)\rangle=|p_1=0\rangle \otimes |p_2=0\rangle \cdots |p_N=0\rangle$, the time-evolved state is obtained as $|\psi(t)\rangle = \mathcal{U}|\psi(0)\rangle$. The numerical simulations are performed using $L$ momentum basis states for each rotor. The resonance condition is incorporated by setting $\hbar_s T=4\pi r/s$, where $r,s\in \mathbb{Z}$ \cite{Santhanam_2022}. In what follows, we fix $\hbar_s=4\pi/T$ and $T=12$. Due to this choice, $U_i^{\rm free}$ becomes an identity operator and does not contribute to the dynamics, and the resulting time-evolution operator is $\mathcal{U}=(U_1^{\rm kick}\otimes U_2^{\rm kick}\otimes \cdots \otimes U_N^{\rm kick}) U_{\rm int}$. 

\begin{figure}[htbt]
	\includegraphics[width=0.99\linewidth]{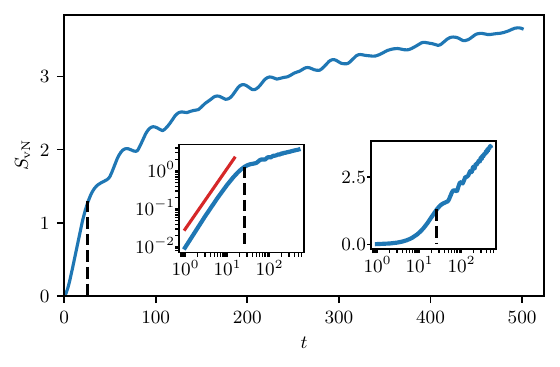}\\
        \includegraphics[width=0.99\linewidth]{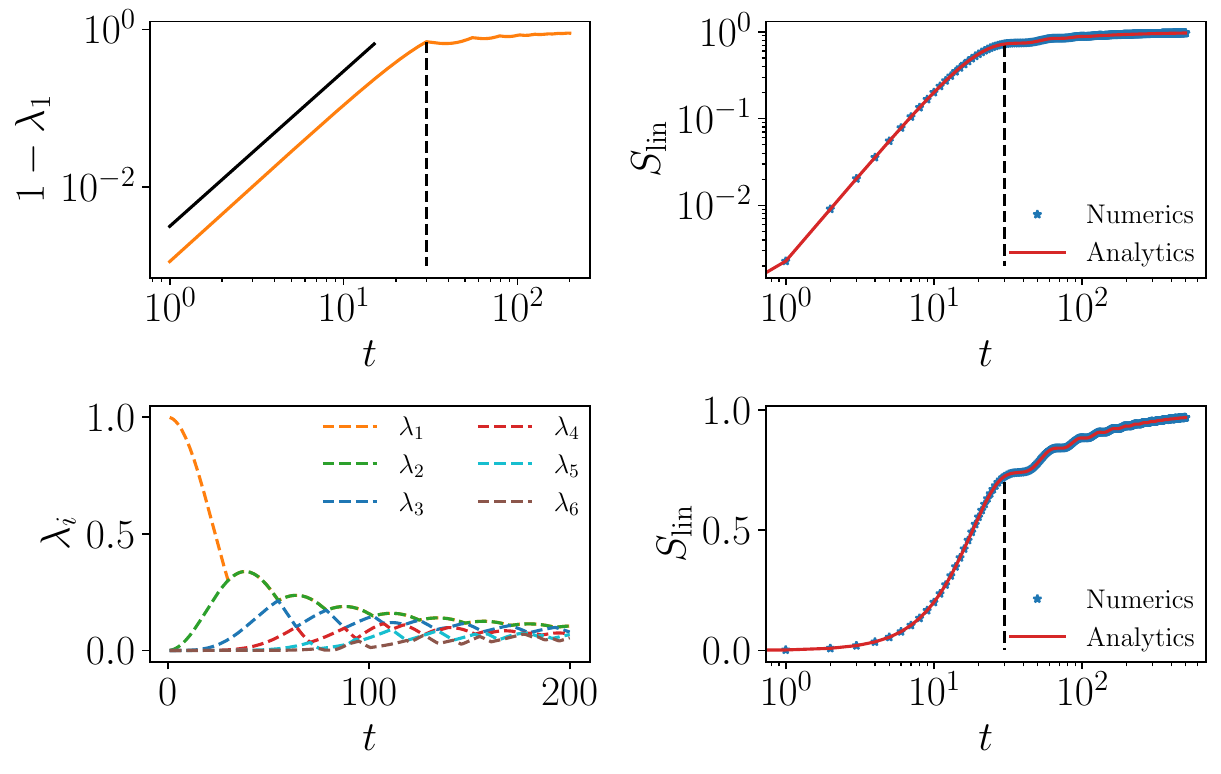}
	\begin{picture}(0,0)
	\put(-83,310){(a)}
	\put(-87,150){(b)}
	\put (38,150){(c)}
	\put(-77,75){(d)}
	\put (38,75){(e)}
	\end{picture}
	\vspace{-0.5cm}
\caption{(a) Evolution of von-Neumann entropy for two-interacting QKR displays two distinct growth profiles: (Left inset) Log-log plot illustrates a super-linear growth, $S_{\rm vN} \sim t^{1.6}$, at initial times and (Right inset) Log-normal plot highlights the logarithmic growth with added oscillation for $t > t^*$. The vertical dashed line represents the crossover time $t^*$ between these two distinct growth profiles. The red solid line in the left inset is a fit to the super-linear growth. (b) illustrates quadratic growth of (1-$\lambda_1$) for $t<t^*$. The black solid line is a fit to this quadratic growth. (c) shows the linear entropy $S_{\rm lin}$ of two-interacting QKR. (d) displays time evolution of the six largest Schmidt eigenvalues $\lambda_i, i=1,2, \dots 6$. (e) displays $S_{\rm lin}$ as log-normal plot to highlight the oscillatory behavior for $t > t^*$. The parameter sets are $K_1=4$, $K_2=5$, $K=0.05$, $\tau_1=1$, $\tau_2=2$, $\Phi_1=0.1$, $\Phi_2=0.15$, $T=12$, and basis size of each rotor, $L=2^{10}$. }
	\label{fig_1}
	\vspace{-.5cm}
\end{figure}
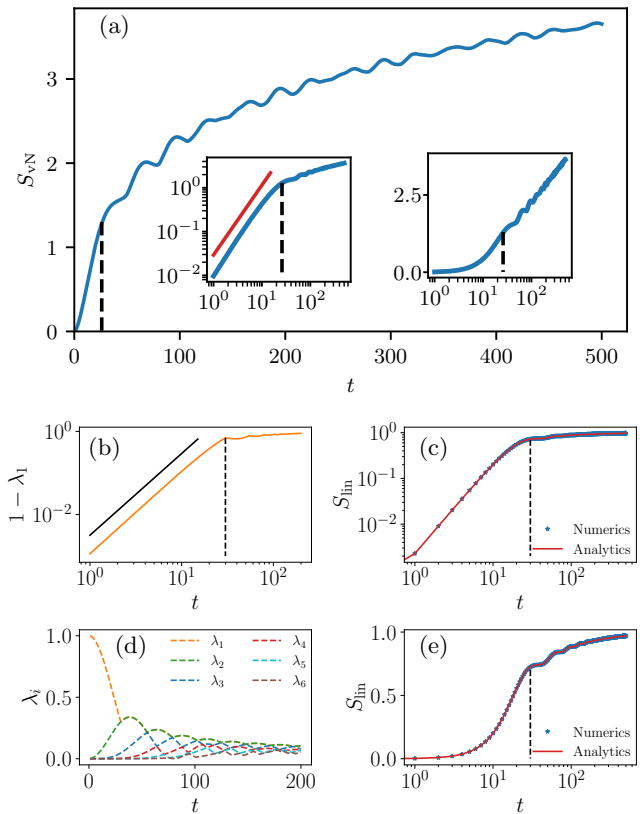
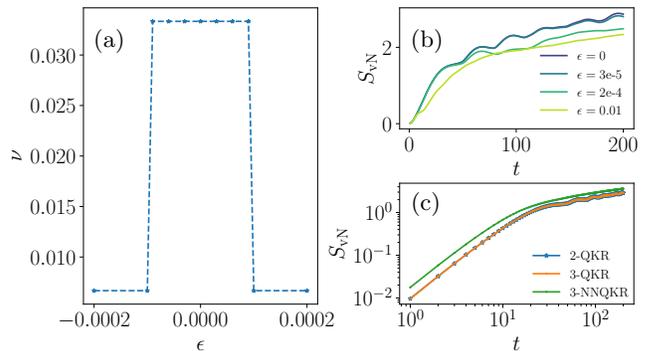
\begin{figure}
	\includegraphics[width=0.995\linewidth]{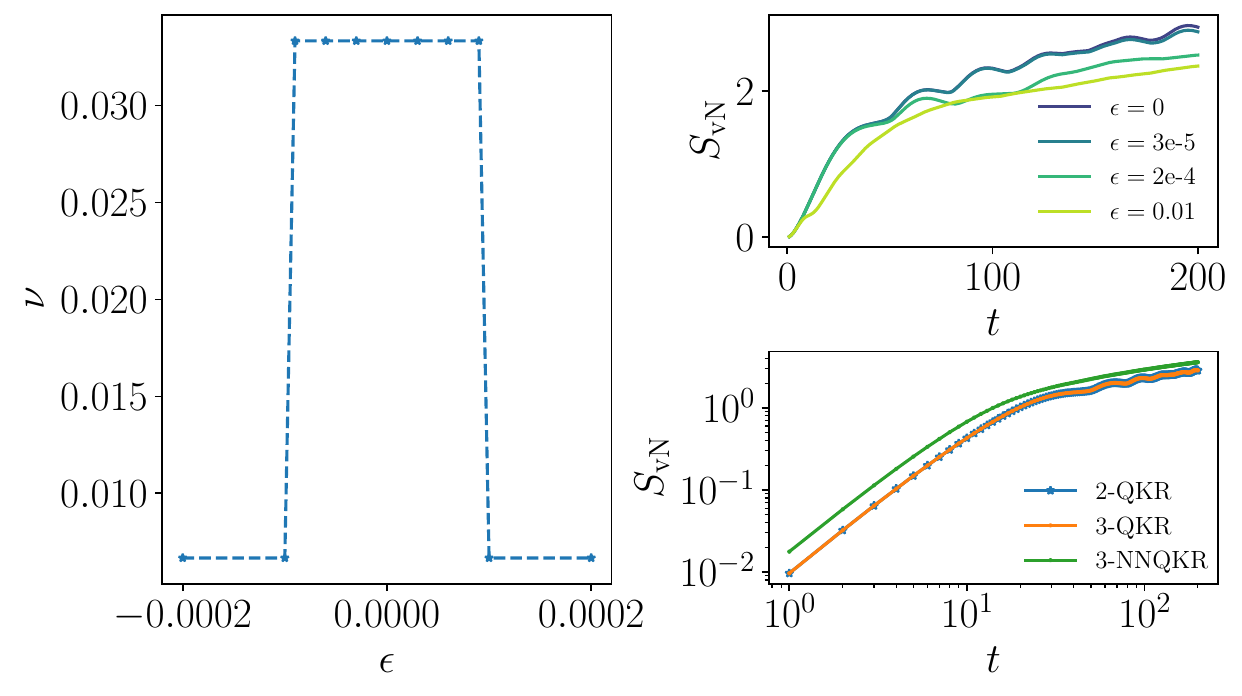}\\
	\begin{picture}(0,0)
	\put(-85,130){(a)}
	\put (35,130){(b)}
	\put (35,68){(c)}
	\end{picture}
	\vspace{-0.5cm}
	\caption{(a) ``Resonance curve'' arising from entanglement dynamics during $t> t^*$ for two interacting QKR. This curve has a high $Q$-factor. The dominant frequency $\nu$ is shown as a function of $\epsilon$. (b) Small deviation from resonance condition significantly changes the growth profile of $S_{\rm vN}(t)$. See text for details. (c) The growth profile of $S_{\rm vN}$ for two- and three-interacting QKR. The kick strength of the third rotor is $K_3=5.5$ and $\Phi_3=0.01$ while the other parameters are the same as in \cref{fig_1} except that $L=2^6$ for both $3$-QKR (all-to-all interacting QKR) and $3$-NNQKR (three nearest-neighbor interacting QKR) in (c).  }
	\label{fig_3}
	\vspace{-.5cm}
\end{figure}

Entanglement is evaluated through von Neumann entropy as $S_{\rm vN}(t)=-\text{Tr} \rho_1(t) \ln \rho_1(t)=-\sum_{j=1}^L \lambda_j(t) \ln \lambda_j(t)$, where $\rho_1(t)$ is the reduced density matrix of the rotor-$1$ at $t$-th kick obtained by tracing out the rest of the system, and $\lambda_j(t)$ are the Schmidt eigenvalues of $\rho_1(t)$. Figure \ref{fig_1}(a) shows the temporal growth of entanglement for two-interacting QKR. It is evident that the entanglement displays two distinct growth profiles. Contrary to the initial linear growth commonly observed in the chaotic regime \cite{Sanku_2020,AnjBhaSan24,Miller_1999}, at resonance entanglement grows super-linearly expressed as $S_{\rm vN}(t)=t^{\mu}$ with $\mu \sim 1.6$ until $t < t^*$; see left inset of \cref{fig_1}(a). However, for $t > t^*$, entanglement displays logarithmic growth with oscillations superimposed on it. Over time, the amplitude of this oscillation diminishes. Furthermore,  oscillations in $S_{\rm vN}(t)$ results from superposition of oscillations observed for $\lambda_j(t) (j=1,2, \dots L)$ (see Fig. \ref{fig_1}(c)). Although these results are for two-interacting QKR, they can be generalized to $N$-interacting QKR. To strengthen this claim, entanglement production for three-interacting QKR ($3$-QKR) is shown in \cref{fig_3}(b). To the best of our knowledge, such distinct growth profiles have never been reported before. 

To gain more insight into distinct growth behaviors of $S_{\rm vN}(t)$, the Schmidt eigenvalues (time dependence suppressed for brevity) $\lambda_1 > \lambda_2 > \dots >\lambda_L \geq 0$ of $\rho_1$ are examined. Figure \ref{fig_1}(b) illustrates the behavior of $1-\lambda_1$ over time, where $\lambda_1$  makes significant contribution to $S_{\rm vN}$. It is evident from \cref{fig_1}(b) that $1-\lambda_1$ initially exhibits a quadratic growth, which at later times $t>t^*$ displays a slower growth accompanied by oscillations. However, with time, the amplitude of oscillation vanishes. The initial quadratic growth of $1-\lambda_1$ can be understood based on Ref. \cite{Arul_2020,Arul_2023}. The initial state of both the rotors can be expressed as a coherent superposition of the respective Floquet states, i.e.,  $|p_i=0\rangle=1/\sqrt{L}\sum_{j=1}^L |\phi_j^i\rangle$, where $|\phi^i\rangle$ is the Floquet state of the $i$-th rotor. For such coherent states, $1-\lambda_1$ is known to display quadratic growth for short times \cite{Arul_2023}. The oscillation observed in $S_{\rm vN}(t)$ arise from that present in individual Schmidt modes -- each $\lambda_i$ has its own distinct frequency and contributes nontrivially to oscillations of $S_{\rm vN}(t)$.

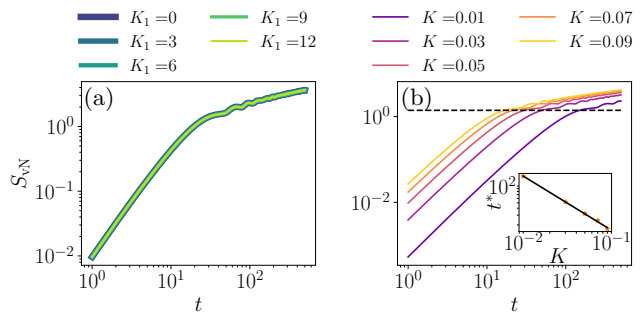
\begin{figure}[htbt]
	\includegraphics[width=0.99\linewidth]{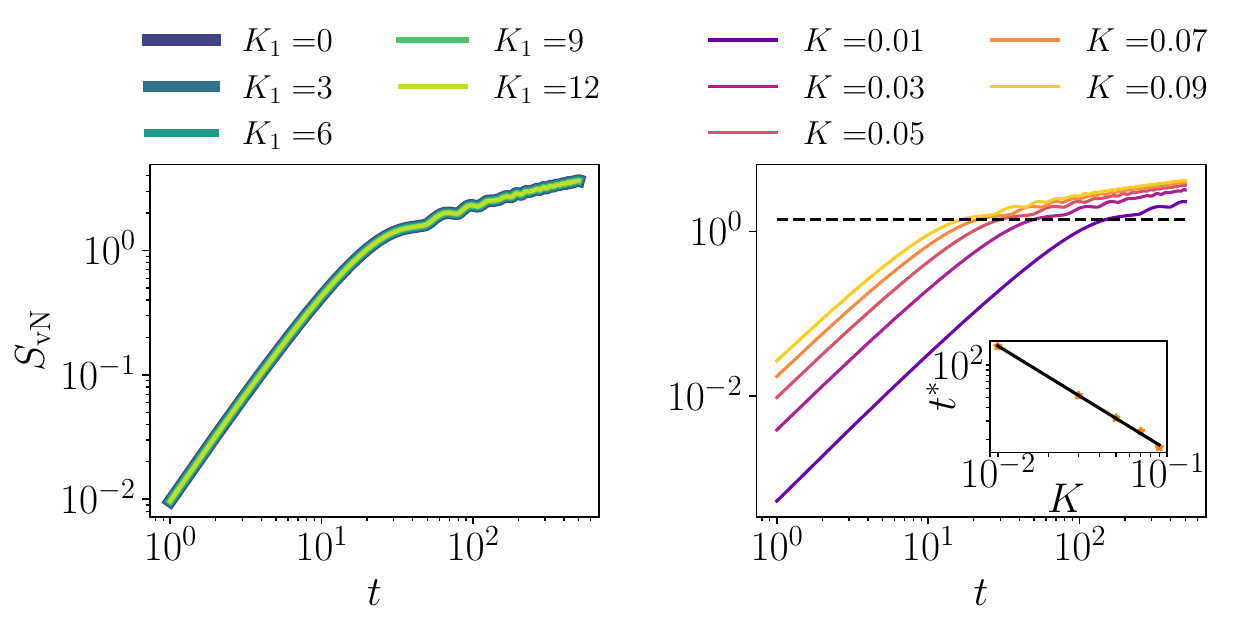}\\
	\begin{picture}(0,0)
	\put(-90,93){(a)}
	\put (30,93){(b)}
	\end{picture}
	\vspace{-0.5cm}
	\caption{The von-Neumann entropy $S_{\rm vN}$ of two interacting QKR for (a) different kicking strengths $K_1$ of rotor-$1$, and  (b) various interaction strength $K$. The growth of $S_{\rm vN}(t)$ is independent of $K_1$ but strongly depends on $K$. Horizontal black dashed line in (b) corresponds to $S_{\rm vN}^*=S_{\rm vN}(t^*)$ after which the growth profile changes showing that $S_{\rm vN}^*$ is independent of $K$. Other parameters are same as in \cref{fig_1}. The log-log plot in the inset of (b) shows that the crossover time $t^*$ has inverse dependence on $K$ through the relation $t^* \sim 1/K$. }
	\label{fig_2}
	\vspace{-.5cm}
\end{figure}

We note that at resonance, as evident in \cref{fig_2}(a), the kicking term in \cref{eq:Ham} does not contribute to entanglement production. This implies that the entanglement growth with $K_i=0$ is identical as that for $K_i \neq 0$. This leads us to a significant conclusion: Two distinct initial states of $i$-th rotor yield the same entanglement dynamics. These are product states ; (a) $|\psi_i(0)\rangle = \sum_{n=1}^L J_n(K_it/\hbar_s) |\phi^i_n\rangle=(U_i^{\rm kick})^t |p_i=0\rangle$, when $K_i\neq 0$ and (b) the zero momentum state $|p_i=0\rangle$ for $K_i=0$. While the initial state in ({\it a}) is a coherent superposition of Floquet states, ({\it b}) is not so. Yet, these distinct type of states display super-linear entanglement growth followed by a logarithmic growth with superimposed oscillations, a feature not observed before and could be useful in quantum resource theory \cite{Alexander_2017}.   

For these reasons, we set $K_i=0$ as it simplifies the analysis. Further, analytical estimates for $S_{\rm vN}(t)$ are also challenging. Therefore, we adopt a simpler quantity, the linear entropy, which is analytically tractable and captures crucial properties of entanglement. The linear entropy is defined as 
\begin{equation}
    S_{\rm lin}(t)= 1-\text{Tr} \rho_1^2(t)   \,.
    \label{eq:slin0}
\end{equation}
Figure \ref{fig_1}(c) displays the growth profile of $S_{\rm lin}(t)$. At short times, for $t<t^*$, $S_{\rm lin}$ grows quadratically with time, \emph{i.e.}, $S_{\rm lin}\sim t^2$, and for $t > t^*$, it grows slowly towards a saturation value of unity. The inset of \cref{fig_1}(c) reveals that $S_{\rm lin}$ also exhibits oscillatory behavior at late times with a amplitude decreasing with time. From \cref{eq:slin0}, it is evident that evaluating $S_{\rm lin}$ crucially relies on determining the purity $\mu_2=\text{Tr} \rho_1^2$. To analytically estimate purity, we perform a coordinate transformation $\Theta = \sum_{i=1}^{N} x_i$ that modifies the interaction term to $U_{\rm int}=\exp[-i K \cos(\Theta)/\hbar_s]$, effectively a ``non-interacting'' single particle kick term. As a consequence, the interactions now appear as a complicated function $\theta(p_1, p_2, \dots p_N, \tau_1,\tau_2,\cdots \tau_N)$ in the momentum space. At resonance, the free evolution term $U^{\rm free}=\exp[-i \theta(p_1, p_2, \dots p_N, \tau_1,\tau_2,\cdots \tau_N) T/2\hbar_s]$ becomes an identity operation, and consequently the many-body QKR effectively reduces to that of a single-particle kicked rotor. Then, purity $\mu_2=\sum_{j=1}^L \lambda_j^2$ becomes equivalent to the participation ratio of a single kicked rotor $\text{PR} = \sum_{p} |c_p|^4$, where the single particle evolved state, $|\psi(t)\rangle=\sum_{p} c_p(t) |p\rangle$. The Schmidt eigenvalues $\lambda_j$ give the probability of finding the particle in Schmidt state $|\zeta_j\rangle$ of the $i$th rotor. For a single kicked rotor at resonance, it can be shown that $\text{PR} =\sum_n J_n(Kt/\hbar_s)^4$, where $J_n(\cdot)$ is the Bessel function of first kind of order $n$. Thus, for the interacting $N$-rotor Hamiltonian in \cref{eq:Ham}, at resonance, the linear entropy can be expressed as 
\begin{equation}  \label{eq:Slin}
    S_{\rm lin} = 1-\sum_n J_n(Kt/\hbar_s)^4 \,.
\end{equation}
Figure \ref{fig_1}(c) illustrates that the analytical result in \cref{eq:Slin} is in excellent agreement with the numerics. Furthermore, \cref{eq:Slin} is not limited to two-interacting QKR but remains valid for any number $N$ of interacting QKR described by the Hamiltonian in \cref{eq:Ham}. Hence, the entanglement evolution reported here is generic for any number of interacting QKR.

How does the entanglement dynamics change with coupling strength $K$ ? Unlike the kicking strength $K_i$, the coupling strength $K$ significantly influences the growth profile of $S_{\rm vN}$. This is evident in \cref{fig_2}. With increasing $K$, numerical simulations in the inset of \cref{fig_2}(b) show that the crossover time $t^*$ from super-linear to logarithmic growth decreases as $t^*\sim 1/K$. This reciprocal relation is in contrast to that observed at off-resonant coupled QKR where linear-to-logarithmic entanglement timescale follows $t^*\sim 1/K^2$ \cite{Sanku_2020}. A plausible argument is that at resonance, the momentum distribution is not of a Gaussian profile, an essential condition required to observe the $1/K^2$ decay of $t^*$. Moreover, lack of analytical expression of the momentum distribution at resonance hinders the analytical calculation of $t^*$. Furthermore, the exponent $\mu$ of the super-linear growth $S_{\rm vN}(t) \sim t^{\mu}$ is independent of the coupling strength as indicated in \cref{fig_2}(b). Additionally, if we denote entanglement at crossover time by $S_{\rm vN}^*\equiv S_{\rm vN}(t^*)$ (horizontal dashed line in \cref{fig_2}(b)), then for $t>t^*$, $S_{\rm vN}(t)$ changes its growth profile and it barely depends on $K$. Thus, the entanglement dynamics is strongly affected by interaction strength rather than the kick strength.

\textit{High quality factor:} Here, we investigate how crucially the oscillatory part of the entanglement growth (for $t > t^*$) depends on the resonance condition. To this end, we define the scaled Planck's constant, $\hbar'_s =4\pi/T'$, where $T'=T+\epsilon$ and $\epsilon \ll T$. As $\hbar'_s$ deviates from the resonance condition, $\hbar_s=4\pi/T$, the dominant frequency $\nu$ present in the power spectrum of $S_{\rm vN}(t)$ (with $t > t^*)$ is tracked as a function of $\epsilon$. 
In \cref{fig_3}, this frequency $\nu$ is shown against $\epsilon$ and the result is reminiscent of a classical ``resonance curve'' of a driven oscillator. It can be noticed that $\nu$ exhibits a sudden drop from from $\nu=0.033$ to nearly zero at a very small $\epsilon=0.0001$. Based on this, the estimated quality factor is high; $Q=\frac{\hbar_s}{\Delta\hbar_s} \approx 10^5$, where 
$\Delta\hbar_s$ denotes the half-width of the resonance curve. Figure \ref{fig_3}(b) shows entanglement growth profile when there is a slight departure from resonance condition. While the short time quadratic growth profile remains akin to that observed at resonance for small $\epsilon$, an absence of oscillations is noticed at later times. This is due to the fact that when $\epsilon>0$, the free evolution part $U_i^{\rm free}$ embedded in $\mathcal{U}$ begins to contribute, resulting in the addition of phases that become significant as time progresses. Thus, even with small $\epsilon \ge 0.0001$ leads the oscillation to vanish immediately. This indicates the system's high quality factor and holds considerable significance in experimental contexts.

\textit{Three-interacting QKR:} To highlight the generic nature of the foregoing results,  $S_{\rm vN}(t)$ is computed for three-interacting QKR. In particular, the entanglement of one rotor (say, rotor-1) with the rest of the system is examined. Remarkably, the results displayed in \cref{fig_3}(c) are identical to that observed for two-interacting QKR. This conclusively establishes the universality of our findings within the context of the interaction considered in \cref{eq:Ham}. Interestingly, \cref{fig_3}(c) also illustrates that for {\sl nearest-neighbor} (NN) interactions of the type $V_{\rm int}=K \sum_i \cos(x_i-x_{i+1})$, the initial super-linear growth of $S_{\rm vN}$ is present, though the amplitude of oscillation in the logarithmic regime dies out. Thus, this demonstrates the generality of super-linear entanglement generation across different interaction potentials under quantum resonance in $N$-interacting QKR.

\textit{Experimental realization:} With current experimental advances in atom-optics kicked rotor setups, the effect of resonance on the quantum dynamics of at least two interacting QKR can be realized using two lasers with incommensurate frequencies \cite{Gadway_2013}. While direct evaluation of entanglement in atom-optics system is still a challenge, it might be possible to infer linear entropy in Eq. \ref{eq:Slin} from the fluorescence images obtained from the evolving atomic cloud \cite{Gadway_19}.

In summary, the ergodic phase of interacting quantum systems are usually associated with linear entanglement production rates. In this work, the all-to-all interacting QKR under the resonance condition displays superlinear entanglement production rate. Two distinct growth profiles can be identified with a crossover at time $t^*$. At short times, for $t < t^*$, the von-Neumann entropy displays super-linear growth, while for  $t > t^*$, growth slows down to a logarithmic form with oscillations superposed on it. The crossover time $t^*$ depends inversely on the coupling strength between the rotors. Further, for the all-to-all interaction considered in \cref{eq:Ham}, {\sl at quantum resonance}, though kicking induces chaotic dynamics it does not contribute to entanglement production. This effectively reduces the many-body problem to that of a single-kicked rotor through a coordinate transformation. As the analytical estimation of von Neumann entropy is challenging, a simpler quantity, namely, linear entropy allows us to obtain analytical insights into the underlying mechanisms driving the observed entanglement growth profiles.  These findings open a new frontier -- faster entanglement production not usually observed in interacting systems. An immediate direction is to extend this investigation for other interaction potential, namely, the point-to-point interaction and also the interplay between resonance and off-resonance conditions on rotors to assess the generality of superlinear entanglement production.

\begin{acknowledgments}
	S.P. would like to thank DST India for the Inspire Faculty Grant and also Michigan State University for the computational facility. JBK and MSS thank I-Hub Quantum Technology Foundation at IISER Pune for partial financial support to this work.
\end{acknowledgments}

\bibliographystyle{apsrev4-1}

\end{document}